\documentstyle[12pt]{article}

\def \bea{\begin{eqnarray}}
\def \eea{\end{eqnarray}}
\def \nn{\nonumber}
\advance \voffset by -.3cm
\advance \hoffset by -.5cm
\begin{document}
\hrule width 0pt
\begin{flushright}
\begin{tabular}{l}
DFPD 97/TH/22 \\
May 1997
\end{tabular}
\end{flushright}
\vspace{1.5cm}
\begin{center}
{\large{\bf
A FERMIOPHOBIC $SU(2)\times SU(2)\times U(1)$ \\
\vskip5pt
 EXTENSION OF THE SM}}
\\
\bigskip
J. Matias \\
Dipartimento di Fisica. \\
 Universit\`a di Padova. \\
Via F. Marzolo 8, I-35131 Padova, Italy
\end{center}

\vspace{2cm}

\begin{abstract}
A new type of gauge extension of the SM is proposed. It is based on a
$SU(2) \times SU(2) \times U(1)$ group with the peculiarity that the
gauge bosons of the extra $SU(2)$ do
not couple to fermions. This feature relaxes some of the
constraints on the masses of the new gauge bosons, leaving the possibility
of having lighter masses than in traditional extensions. The model exhibits
several interesting properties, it is anomaly free and at tree level
it does not have FCNC while loop induced effects are strongly suppressed.
Also,
from the analysis of $\Delta \rho$ at one loop, two
configurations for the vevs giving rise to a screening
phenomenon are identified.
One of these configurations
 can also be related with the
Bess model. A tree level fit to the most recent electroweak data
is performed confirming the possibility of having new light gauge
boson masses.
The constraints coming from different FCNC processes like
$b \rightarrow s \gamma$, $B_{0}-{\bar B}_{0}$ and $K_{0}-{\bar K}_{0}$
mixing are also taken into account. Finally, a generalization of this
model for the case of having several extra copies of  $SU(2)$ groups is
commented,
focusing on the presence of screening configurations
and the corresponding mass spectrum.
\end{abstract}

\vspace{1cm}

\begin{center}
{
Contribution to the XXXIInd Rencontres de Moriond: \\
''Electroweak Interactions and Unified Theories'', Les Arcs, France, March 1997}
\end{center}

\newpage

In the post-LEPI era
the extraordinary success of the Standard Model (SM) in explaining data
has been confirmed. As a consequence any acceptable alternative should
be able to repeat its success while allowing for new physics at a not
very high scale to be interesting for present and future colliders.
Extensions of the SM based on enlarging the gauge group with an extra
$SU(2)$ could fulfill this requirement. However, in most of
these models (Standard
Left-Right (LR)$^{1)}$ or ununified models$^{2)}$)
the lower bounds on the new gauge boson masses are still quite high and
new problems like anomalies or Flavour Changing Neutral Currents
(FCNC) appear.

Instead, one can think of a particular type of $SU(2)$ extension of the
SM that we have called `fermiophobic'$^{3)}$, which seems to naturally
satisfy all existing constraints and, at the same time,  avoids most of
the problems of the other models while allowing for light new gauge
bosons.

This model, defined by the transformation properties of fermions
under the gauge group $SU(2)_{L}\times SU(2)_{R}\times U(1)_{\tilde Y}$,
\bea
q_{L}&=&\left( \begin{array}{c} u_{L} \\ d_{L} \end{array} \right) \sim
(2,1,1/6) \quad \quad
l_{L}=\left( \begin{array}{c} \nu_{L} \\ e_{L} \end{array} \right)
\sim (2,1,-1/2) \nn \\
u_{R} &\sim& (1,1,2/3) \quad \quad d_{R} \sim (1,1,-1/3)
\quad \quad
 e_{R} \sim (1,1,-1) \nn
\eea
is automatically free of gauge anomalies. Contrary to standard LR
models, it does not require the existence of right-handed
 neutrinos.

The field content of the scalar
sector consists of two doublets and one bidoublet
\bea
\phi_{L}=\left( \begin{array}{c} \phi_{L}^{0} \\ \phi_{L}^{-}
\end{array} \right) \sim (2,1,-1/2) \quad
\phi_{R}=\left( \begin{array}{c} \phi_{R}^{0} \\ \phi_{R}^{-}
\end{array} \right) \sim (1,2,-1/2) \quad
\phi_{LR}=\left( \begin{array}{c} \phi_{1}^{0}\,\,\,\, \phi_{2}^{+} \\
\phi_{1}^{-}\,\,\,\, \phi_{2}^{0} \end{array} \right) \sim (2,2,0)  \nn
\eea
 The  absence of  right-handed neutrinos makes unnecessary the use of
scalar triplets.
If we assume that there are no new sources of CP-violation, the
previous sixteen scalar degrees of freedom organize in a set of eight
charged scalars: four Goldstone bosons ($G^{\pm}$, $G^{\prime
\pm}$)
and four charged physical scalars $H_{1}^{\pm}$, $H_{2}^{\pm}$ with
mixing angle $\beta_{\pm}$. And eight neutrals: four scalars
($H_{3}^{0}$, $H_{4}^{0}$, $H_{5}^{0}$ and $H_{6}^{0}$), one of them can
be identified with the SM-Higgs
like, and four pseudoscalars, two Goldstone bosons $G^{0},G^{\prime
0}$ and two physical pseudoscalars $H_{1}^{0},H_{2}^{0}$ with mixing angle
$\beta_{0}$.

Finally, concerning the gauge sector we have four charged gauge
bosons ($W^{\pm}$, $W^{\prime \pm}$) with
mixing angle $\alpha_{\pm}$, the photon $\gamma$ and two neutrals
($Z$,
$Z^{\prime}$) with mixing angle $\alpha_{0}$.
Moreover, the gauge coupling constants $g_{L}$ and $g_{R}$
 can be different and its
ratio $g_{R}/g_{L}$ is called $x$.

Given these quantum number assignments and assuming a pattern of
real vevs for the scalar fields
\bea
<\phi_{L}>={1 \over \sqrt{2}}\left( \begin{array}{c} v_{L} \\ 0
\end{array} \right) \quad
<\phi_{R}>={1 \over \sqrt{2}}\left( \begin{array}{c} v_{R} \\ 0
\end{array} \right)  \quad
<\phi_{LR}>={1 \over \sqrt{2}}\left( \begin{array}{c} v_{1}\,\,\,\, 0 \\
0\,\,\,\, v_{2} \end{array} \right),\nn \eea
 one can automatically write down the
lagrangian and derive the gauge boson masses (both
doublets and the bidoublet contribute)
 and the relevant vertices $^{3)}$.
Notice that the Yukawa part of the lagrangian
turns out to be as in the SM, only the $\phi_{L}$ doublet can
couple to fermions,
avoiding automatically the tree level FCNC
problems of standard LR models. Moreover,
at the level of gauge interactions of fermions an important remark
 is the strong suppression of the $W^{\prime
\pm}$ (mass eigenstate)
coupling to
fermions, while the corresponding $Z^{\prime}$ coupling is suppressed
only in the large $g_{R}$ limit.

We have imposed a theoretical constraint to the
model at the level of radiative corrections.
 We require that all  contributions quadratic in
the masses of the scalars   cancel in $\Delta \rho$ at
one-loop.
The implications of such {\it screening} requirement
however, go well beyond to the mere cancellation of quadratic
contributions. From the explicit expression of $\Delta \rho^{4)}$
one sees that the fulfilment
of that condition can be reached in  different ways.
Two of them translate into constraints on the vevs, $\beta$
angles and scalar masses:

i) $v_{R}\rightarrow \infty $ and $\beta_{\pm}=\beta_{0}$.
$\rho$ at tree level is exactly one and the one-loop correction
corresponds to the one of a two doublet model.
 However, the new gauge boson masses become infinite.

ii) $v_{R}=v_{L}/x$, $v_{1}=v_{2}=v$, $\beta_{\pm}=\beta_{0}$ and
${\rm tan} \beta=-M_{W^{\prime}}/M_{W}$. When the mass of the
$H_{1}^{+}$ is degenerate with $H_{5}^{0}$ and $H_{2}^{+}$ with
$H_{2}^{0}$ the one-loop quadratic contributions to $\Delta \rho$
cancel. Masses and mixing angles  become
extremely simple in that case,
\bea
M_{W}^{2}={g_{L}^{2} v_{L}^{2} \over 4} \quad \quad
M_{Z}^{2}={g_{L}^{2} v_{L}^{2} \over 4} {x^{2} \over
x^{2} -s_{W}^{2} x^{2} -s_{W}^{2}} \quad \quad
M_{W^{\prime}}^{2}=M_{Z^{\prime}}^{2}={g_{L}^{2} \over 4}({v_{L}^{2} + 2
v^{2} (1+x^{2})})  \nn
\eea
and the mixing angles are
$ \tan \alpha_{\pm}=1/x $ and  $ \tan \alpha_{0}=1/\sqrt{x^{2} -
s_{W}^{2} x^{2} - s_{W}^{2}} $
with $s_{W}$ the sinus of the Weinberg angle. Notice from the
previous expressions that the new gauge bosons should be
degenerate with a mass always heavier than $M_{W}$. One observes, also, that
the theory requires  $x$ to be large in order to have a
small tree level $\rho$ parameter. Moreover, an interesting property
of that configuration is that when one imposes
a further restriction on the vevs  ($v_{R}=\sqrt{2} v$) the
mass predictions of our model coincide with those of the Bess
model$^{5)}$ in a restricted 4-parameter space.

One of the most important experimental constraints on the model comes
from a tree level fit to the electroweak data at
the Z peak,
the low energy data from neutrino-hadron scattering and atomic parity
violation experiments (fourteen observables in total)$^{6)}$. The model
has, concerning
only bosonic interactions, seven parameters ($g_{L}$, $g_{R}$, ${\tilde
g}$, $v_{R}$, $v_{L}$, $v_{1}$, $v_{2}$) that we have translated in
terms of the three input quantities of the SM precision tests ($G_{F}$,
$\alpha$, $M_{Z}$) and four extra input parameters: $\alpha_{\pm}$,
$\alpha_{0}$, $M_{W^{\prime}}$ and $x$. The fit was done by
adding to the SM predictions the deviations due to the model. These
deviations depend on the four extra input parameters. A
$\chi^{2}$ minimization procedure is used to determine their best
values.

The results of the fit are:

a) in the general case the fit turns out to be quite insensitive to
$M_{W^{\prime}}$ for fixed values of $x$. So we kept $x$ and
$M_{W^{\prime}}$ fixed in
a range $1\leq x \leq 15$ and $100\leq M_{W^{\prime}}\leq 1000$ leaving
$\alpha_{\pm}$ and $\alpha_{0}$ free. In that scenario the experimental data
allows a mixing angle $\alpha_{\pm}, \alpha_{0}$ in the range
$10^{-2}-10^{-3}$ and larger mixing angles are allowed for larger $x$
values. The masses of the new gauge bosons $W^{\prime}$, $Z^{\prime}$
can be in a broad range allowing also for light values, even of the
order of 150 GeV if $x$ is sufficiently large ($x>5$).

b) in the particular screening configuration ii) mentioned above, we no
more have four but just two free parameters ($x$ and $M_{W^{\prime}}$).
The fit prefers large values of $x$ approaching in that limit the SM and
becoming then insensitive to the value of $M_{W^{\prime}}$.

Other bounds on the masses of the new gauge bosons and
the charged
Higgs bosons could come from one-loop FCNC processes. While they are
quite
severe in standard LR models$^{7)}$, in our model they become
naturally relaxed due to the absence of coupling between
the $\phi_{R}$, the
$\phi_{LR}$ and
the $W_{R}^{\pm}$ field (in the interacting basis)
 with ordinary fermions. The $b\rightarrow s \gamma$,
$B_{0}-{\bar B}_{0}$ and $K_{0}-{\bar K}_{0}$ mixing have been
examined finding an exclusion region for light values of the mass of the
charged Higgs bosons (we take them to be degenerate) when small values of $x$
are taken. However, no sensitivity to $M_{W^{\prime}}$ was found.
In the
$b\rightarrow s \gamma$ process, for instance, this is due to the
strong suppression of the first correction to the
$W^{\pm}$ boson exchange amplitude that, in our model, is of
order $\alpha_{\pm}^{2}$
while in standard LR models  is of order $\alpha_{\pm}$.

 Bounds like those coming from neutrinoless
double beta decay are, obviously, innocuous to our model.

The study of the limits from the Tevatron, and, of the possible signals of
the new gauge boson particles in present
and future colliders  is presently being
completed$^{3)}$.

Finally the existence of such  screening configurations has been
studied also in a more general case with additional $SU(2)$
groups$^{4)}$. The result, again positive, is particularly smart
since it allows to obtain the exact gauge mass spectrum of an $SU(2)
\times SU(2)^{n} \times U(1)$ theory starting just from the information
of the screening of quadratic contributions to $\Delta \rho$. The
pattern of new gauge boson masses is similar to that of the case ii).
Each pair of new gauge bosons masses is degenerate and
differs
from the other pairs. $M_{W}$ is exactly the same as in the
case $n=1$ and the $M_{Z}$
mass is a straightforward generalization of the $n=1$ case.

The work reported here was done in collaboration with A. Donini, F.
Feruglio and F. Zwirner.

\bigskip

\noindent{\bf \large References}
\bigskip

\noindent [1] {\sc J.C. Pati and A. Salam}, {\sl Phys. Rev.} {\bf D10}
(1974) 275; {\sc R.N. Mohapatra and J.C. Pati}, {\sl Phys. Rev.}
 {\bf D11} (1975) 566 and 2558; {\sc G. Senjanovic and R.N. Mohapatra}, {\sl
Phys. Rev.} {\bf D12} (1975) 1502.
\medskip

\noindent [2] {\sc H. Georgi, E.E. Jenkins and E.H. Simmons}, {\sl Nucl. Phys.}
{\bf B331} (1990) 541.
\medskip

\noindent [3] {\sc A. Donini, F. Feruglio, J. Matias and F. Zwirner},
"Phenomenological aspects of a  fermiophobic
$SU(2)\times SU(2)\times U(1)$
extension of the SM", Padova/Cern Prep. DFPD 97/TH/23 1997.
\medskip

\noindent [4] {\sc J. Matias and A. Vicini},
"The $\rho$ parameter and the screening phenomena for sequential W and Z
gauge bosons",
Padova/Desy Prep. 1997.
\medskip

\noindent [5] {\sc R. Casalbuoni,S. de Curtis,D. Dominici and
R. Gatto}, {\sl Nucl. Phys.} {\bf B282} (1987) 235.
\medskip

\noindent [6] {\sc Particle Data Group}, {\sl Phys. Rev.} {\bf D54} (1996) 1;
The LEP Electroweak Working Group; prep. CERN-PPE/96-183 and Internal
Note LEPEWWG 97-01;  {\sc C.S.Wood et al.} {\sl Science} {\bf 275} (1997) 1759.
\medskip

\noindent [7] {\sc P.Langacker and S.Uma Sankar}, {\sl Phys. Rev.}
{\bf D40} (1989) 1569.

\end{document}